\documentclass[prd,aps,showpacs,nofootinbib,floatfix,onecolumn,superscriptaddress]{revtex4-2}
\usepackage{slashed}
\usepackage{float}
\usepackage{amsmath,graphicx,color,epsfig,ulem}
\usepackage{bm}
\usepackage{cancel}
\definecolor{purple}{rgb}{0.5,0,0.5}
\usepackage[colorlinks=true, pdfstartview=FitV, linkcolor=purple, citecolor=blue, urlcolor=blue]{hyperref}

\newcommand{\beq}{\begin{equation}}
\newcommand{\eeq}{\end{equation}}

\begin{document}

\title{Photon Gas Thermodynamics with $\kappa$-generalized statistics at Planck-Scale}
\author{Xinyi Yang}
\email{estelle0318@163.com}
\affiliation{Department of Physics, Shanghai Normal University, Shanghai 200234, People's Republic of China}

\author{Qi Xiong}
\email{xiongqi@bao.ac.cn}
\affiliation{National Astronomical Observatories, Chinese Academy of Sciences, 20A Datun Road, Beijing 100101, 
People's Republic of China}

\author{Guifeng Su}
\email{guifeng_su@shnu.edu.cn}
\affiliation{Department of Physics, Shanghai Normal University, Shanghai 200234, People's Republic of China}

\author{Yi Zhang}
\email[Corresponding author: ]{yizhang@shnu.edu.cn}
\affiliation{Department of Physics, Shanghai Normal University, Shanghai 200234, People's Republic of China}

\begin{abstract}
Kaniadakis, or $\kappa$-generalized, statistical mechanics provides a flexible framework for describing 
complex systems in the relativistic regime and predicts a richer phenomenology compared to its standard 
Maxwell-Boltzmann counterpart. In the present work, we extend the investigation of photon gas thermodynamics 
at the Planck scale within doubly special relativity to the framework of $\kappa$-generalized 
statistical mechanics. By adopting the $\kappa$-generalized statistical approach, we derive the principal 
thermodynamic quantities of a photon gas at the Planck scale, including the internal energy $U$, Helmholtz 
free energy $F$, pressure $P$, entropy $S$, and heat capacity $C_{\rm V}$. We find that these 
$\kappa$-deformed thermodynamic quantities exhibit nontrivial dependence on the deformation parameter 
$\kappa$. We then numerically evaluate these quantities as functions of temperature $T$, and reveal that, 
as $T$ approaches the Planck scale, the $\kappa$-deformed thermodynamic quantities are significantly 
suppressed relative to those in special relativity, while in the low-temperature regime they are 
enhanced and exceed their counterparts in special relativity, as a consequence of the $\kappa$-generalized 
statistics.
\end{abstract}

\maketitle

\section{Introduction}

Quantum gravity aims to unify general relativity with quantum mechanics at the Planck scale. To date, 
however, a fully consistent theory of quantum gravity remains absent. Numerous approaches to quantum gravity 
phenomenology suggest the existence of a fundamental minimal length scale (or equivalently, a maximum energy 
scale)~\cite{Ameli2001,Glik2001,Ameli2001,Glik2002,Ameli2002,Mague2002,Mague2003,Mague2004}. 
The Planck length, constructed from the gravitational constant $G$, the quantum of action $\hbar$, and the 
speed of light $c$: $l_{\rm P} = \sqrt{\hbar G/c^3} \sim 10^{-35}$ m, is widely regarded as the most natural 
candidate for such a scale. Its inverse is the Planck energy scale, 
$E_{\rm P} = l_{\rm P}^{-1} = \sqrt{\hbar c^5/G} \sim 10^{19}$ GeV. 
Beyond these scales, the classical description of spacetime breaks down, and quantum gravitational effects 
are expected to become significant. However, the Planck scale also introduces a series of paradoxes. For 
instance, with respect to which reference frame is the Planck length defined? This raises a conflict with 
the equivalence principle, as observers in different inertial frames would measure different values of 
$l_{\rm P}$ due to Lorentz-Fitzgerald contraction. These apparent puzzles are deeply tied to the fundamental 
understanding of the principle of relativity, the very cornerstone of special relativity (SR).

One type of solutions falls under the name doubly special relativity (DSR), in which the minimal length 
$l_{\rm P}$ remains invariant under spacetime transformations. The first such model was proposed by 
Amelino-Camelia~\cite{Ameli2001} (DSR1). Later, another model (DSR2) was introduced by 
Amelino-Camelia~\cite{Ameli2002} 
and independently by Magueijo and Smolin~\cite{Mague2002,Mague2003} (We refer to the latter as the MS 
model). For recent discussions and reviews, see Ref.~\cite{Ameli2010} and references therein. In these 
models, there are two invariant quantities: the speed of light $c$ and an upper limit of energy scale 
$\Lambda = E_{\rm P}$, which are postulated to be observer-independent. 
To consistently incorporate this second invariant alongside the other principles of SR, the well-known 
dispersion relation (or mass-shell condition) $E^2 - p^2 = m^2$ must be modified to 
$E^2 - p^2 = m^2 \left(1 -\frac{E}{\Lambda}\right)^2$,
where $E$ and $p$ are the energy and magnitude of the three-momentum of the particle, respectively, and 
$m$ is its mass. 
Consequently, the integration over energy-momentum space is also modified~\cite{Ameli2000,Ameli2009}.
These modifications lead to further investigations and/or various predictions, including black 
hole evaporation~\cite{Sales2009}, cosmology~\cite{Alex2003}, 
cosmic ray spectra~\cite{Ameli2003}, $\gamma$-ray bursts~\cite{Ameli2002,Gamb1998,Ellis2000}, neutrino 
propagation~\cite{Alfaro1999}, dark energy~\cite{Mersini2001}, and astrophysical implications~\cite{Kar2026}. 
For a recent review of quantum spacetime phenomenology, see Ref.~\cite{Ameli2013}.
Some interesting DSR effects on the thermodynamic properties of a photon gas at the Planck scale have 
also been studied in Ref.-s.~\cite{Cama2007,Das2010}, which is the topic the present paper aims to address.

On the other hand, it has been increasingly recognized that the conventional Maxwell-Boltzmann (MB) 
distribution, derived from Boltzmann-Gibbs-Shannon (BGS) entropy, exhibits substantial limitations when 
applied to a broad class of systems, including those characterized by nonequilibrium dynamics, 
long-range interactions, and thermal fluctuations~\cite{Yalcin2018}, as well as those involving highly 
relativistic regimes~\cite{Hase1985}. In particular, the MB distribution is unable to account for the 
non-exponential, power-law asymptotic behavior observed in the ultra-relativistic cosmic ray spectrum 
over an exceptionally wide range of particle fluxes~\cite{Bierm2001}. These empirical shortcomings 
provide a strong motivation for the introduction of a more general statistical framework in which the 
MB distribution emerges as a specific limiting case.

Several extensions of the BGS entropy have been proposed in recent years, arising from various 
generalizations of the Boltzmann-Gibbs (BG) statistical mechanics. In its standard form, the entropy 
of a system is defined as  
$S= -k_{\rm B} \sum_{i} p_i \ln p_i$, where $k_{\rm B}$ is Boltzmann constant, and $p_i$ denotes the 
probability of occupying a given microstate.
This classical formulation has been modified based on distinct physical motivations: non-extensive 
statistical mechanics leads to the Tsallis entropy~\cite{Tsallis1988,Tsallis1998}, 
quantum-gravitational modifications give rise to the Barrow entropy~\cite{Barrow2020}, 
and information-theoretic approaches yield the R\'{e}nyi~\cite{Reny1959} and 
Sharma-Mittal~\cite{Sharma1977} entropies. 
Predictions of these models have been tested in various physical contexts, see, e.g. 
Ref.-s~\cite{Shaba2020,Luci2021a,Luci2021b,Jizba2022,Tava2018,Sari2018,Sari2020}, to name a few. 
A particularly intriguing relativistic generalization of BGS entropy, inspired by Lorentz symmetries, 
is the 
Kaniadakis entropy~\cite{Kania2001a,Kania2001b,Kania2002a,Kania2002b,Kania2005a,Kania2005b,Kania2006} 
based on the modified entropy formula $S_{\kappa} = -k_{\rm B} \sum_{i} f_i \ln_{\kappa} f_i$, 
where $\ln_{\kappa}$ is $\kappa$-deformed logarithm, which was introduced 
in Ref.~\cite{Kania2001a,Kania2001b}, and $f_i$ is the generalized Boltzmann distribution for the 
$i$-th microstate (see Sec. III for details). Note that the ordinary logarithmic and exponential 
functions, and thus conventional statistical mechanics, are readily recovered in the $\kappa \to 0$ 
limit. The appearance of Kaniadakis entropy within the SR context enables a reexamination of 
relativistic thermodynamic problems, lending support to the Einstein-Planck conjecture 
that a moving frame experiences a reduction in relativistic temperature.

Kaniadakis entropy, along with the associated $\kappa$-generalized statistical mechanics, constitutes 
a flexible framework that may potentially accommodate the diversity of relativistic physical contexts 
in which the MB distribution proves inadequate. Indeed, $\kappa$-generalized statistical mechanics 
preserve important features of classical statistical mechanics on the one hand, on the other hand, 
$\kappa$-generalized statistical mechanics lead to a power law tail in the associated statistical 
distribution in relativistic regime, while the ordinary statistical mechanics leads to exponential 
tails. This formalism has found applications in 
gravity and cosmology~\cite{Morad2020,Lymp2021,Drep2022,Hern2020} (see also Ref.~\cite{Luci2022} for 
a recent review), quantum entropy\cite{Sant2012,Oura2015}, sociophysics and network 
theory~\cite{Maced2013,Stell2014}, nuclear physics~\cite{Tewel2003,Shen2018}, as well as in black 
hole thermodynamics~\cite{Cimid2023}. 
Further applications have 
been explored in plasma physics, astrophysics, information theory, and other disciplines, see 
Ref.~\cite{Kania2021} for a comprehensive overview and references therein. Nevertheless, to the best 
of our knowledge, a detailed investigation of photon gas 
thermodynamics at the Planck scale, within the framework of 
$\kappa$-generalized statistical mechanics has not yet been carried out.

Building on the foregoing premises, in this paper we investigate the thermodynamics of a photon gas 
within the DSR framework, adopting the perspective of $\kappa$-generalized statistical mechanics. 
The principal thermodynamics quantities of a photon gas are derived, upon employing both the modified 
dispersion relation characteristic of the MS model and the $\kappa$-deformed statistical distribution. 
Specifically, we numerically evaluate these thermodynamic quantities of the photon gas in the regime 
of Planck scale.

The remainder of the paper is organized as follows. Sec. II provides a concise overview of the DSR 
framework and presents the modified dispersion relation relevant to a photon gas. In Sec. III, we 
outline the general theoretical formalism of $\kappa$-generalized statistical mechanics. Sec. IV is 
devoted to the derivation and numerical analysis of the main $\kappa$-deformed thermodynamic 
quantities of a photon gas, including internal energy, Helmholtz free energy, heat capacity, 
radiation pressure, and entropy. Moreover, we compare these quantities with those obtained from 
conventional statistical mechanics within both SR and DSR contexts. Finally, in Sec. V, the 
summary and some concluding remarks are reported.

\section{DSR and Modified Disperse Relation}
\label{sec:DSR}

Among the various proposals for realizing DSR, the MS scheme due to Magueijo and Smolin~\cite{Mague2002,Mague2003} 
stands out as particularly attractive. In this formulation, the Lorentz algebra is left intact, yet its action on 
phase space is realized nonlinearly, thereby naturally introducing the invariant energy scale $\Lambda$. 
The smooth limit $\Lambda \to \infty$ consistently recovers standard SR.

According to the MS model of DSR, a general isotropic modified dispersion relation can be written in the following 
form:
\beq
 E^2 f_1^2(E/\Lambda) - p^2 f_2^2 (E/\Lambda) = m^2 ~,
\eeq
where $f_1$ and $f_2$ are two to be determined functions depending on energy $E$ and $\Lambda$, respectively,
and $\Lambda$ is the Planck energy scale. 
In current work, we take 
\beq
 f_1^2= f^2_2 = \left(1-\frac{E}{\Lambda}\right)^{-2} ~.
\eeq

The modified dispersion relation we adopt is given in \cite{Mague2002,Mague2003} as
\beq
\label{eq:mdisp}
 E^2 - p^2 = m^2 \left(1 -\frac{E}{\Lambda}\right)^2 ~.
\eeq 
It is worth noting that the parameter $m$ appearing in Eq.~\eqref{eq:mdisp} is interpreted as an invariant 
mass, as it remains unchanged under DSR transformations. This contrasts with the case of SR, where the dispersion 
relation is $E^2=p^2+m^2$, and $m$ corresponds directly to the rest mass. 
The physical rest mass energy $m_0$ is obtained by setting $p=0$ in Eq.~\eqref{eq:mdisp}, which yields
\beq
 m_0 = \frac{m}{1+m/\Lambda} ~.
\eeq
Consequently, within the MS framework, the physical states are restricted to energies satisfying 
$0 < E < \Lambda$.
  
We emphasize that the choice of this dispersion relation is not unique, only experiments can determine 
the correct dispersion relation in nature. Furthermore, a modified dispersion relation does not necessarily 
imply an energy-dependent speed of light~\cite{Hoss2006}. In the MS model, the dispersion relation for 
photons is the same as in SR, and the speed of light $c$ remains an invariant quantity. However, there 
exists a finite upper limit of the photon energy $\Lambda$. 

\section{$\kappa$-generalized Statistical Mechanics}
\label{sec:kapsm}

As mentioned in the Introduction, conventional statistical mechanics exhibits certain limitations when 
applied to the relativistic systems. In highly relativistic physical environments, the standard 
exponential function decays too rapidly in the asymptotic tail region, thereby failing to accurately 
capture the long-tail and power-law behaviors frequently encountered in high-energy phenomena. 
To address these theoretical 
shortcomings, Kaniadakis introduced the so-called Kaniadakis entropy~\cite{Kania2001a,Kania2001b},
\beq
\label{eq:KS}
 S_{\kappa} = - k_{\rm B} \sum_{i} f_i \ln_{\kappa} f_i ~, 
\eeq
where the $\kappa$-deformed logarithm function, $\ln_{\kappa}$, is defined by
\beq
\label{eq:kap_ln}
 \ln_{\kappa}(x) = \frac{x^\kappa - x^{-\kappa}}{2\kappa} \, ,
\eeq
in which the dimensionless deformation parameter $|\kappa|< 1$. The generalized Boltzmann distribution 
for the $i$-th microstate, $f_i$, is,
\beq
\label{eq:fki}
 f_i\,=\, \alpha_{\kappa} \exp_{\kappa} 
 \left[-\beta \left(E_i-\mu\right)/\sqrt{1-\kappa^2} \right] \, ,
\eeq 
where $\alpha_{\kappa} = \left(\frac{1-\kappa}{1+\kappa}\right)^{1/2\kappa}$, 
$\beta = 1/k_{\rm B} T$,
and $T$ and $\mu$ being the temperature and chemical potential of the system, respectively. 
The $\kappa$-deformed exponential function $\exp_{\kappa}(x)$ is defined as:
\beq
\label{eq:kap_exp}
 \exp_{\kappa}(x) = \left( \sqrt{1+\kappa^2 x^2} + \kappa x \right)^{1/\kappa} \, .
\eeq
This definition guarantees the crucial calculus property 
$\frac{d}{d_{\kappa}x} \exp_{\kappa}(x) = \exp_{\kappa}(x)$.
The deformed Kaniadakis exponential is a positive monotonic increasing function for $x \in R$.
Asymptotically, the $\exp_{\kappa}(x)$ follows the power law: 
$\exp_{\kappa} (x) \sim |\kappa x|^{\pm 1/|\kappa|}$, as $x \to \pm \infty$. In addition, 
it is required to satisfy 
the self-dual condition~\cite{Kania2001a}
\beq
 \exp_{\kappa}(x) \, \exp_{\kappa}(-x) = 1 ~.
\eeq
which ensures that the deformation remains compatible with the symmetry of the 
standard exponential function. 

In the $\kappa \to 0$ limit, the $\kappa$-deformed exponential and logarithmic 
functions reduce to the ordinary exponential and logarithmic functions, respectively, i.e.,
\beq
 \lim_{\kappa \to 0}\exp_{\kappa}(x) = \exp(x) ~, \quad \lim_{\kappa \to 0}\ln_{\kappa}(x) = \ln(x) ~.
\eeq 

Physically, the deformation parameter $\kappa$ quantifies the deviation from the MB distribution. In 
the $\kappa = 0$ limit, the ordinary exponential and logarithmic functions, together with the BGS 
entropy and the MB distribution, are fully recovered. For $\kappa \neq 0$, the deformed exponential 
exhibits power-law asymptotic behavior, in contrast to the purely exponential decay characteristic 
of the standard statistical mechanics. 

The $\kappa$-generalized statistical mechanics framework inherently preserves the fundamental 
relativistic symmetries. In the limit of $\kappa \to 0$, the conventional formulation of MB 
statistical mechanics is consistently recovered, including the well-known Maxwell–J\"{u}ttner (MJ) 
statistical distribution~\cite{Jutt1911}, which constitutes the relativistic generalization of the 
non-relativistic MB distribution. In other words, the conventional relativistic MB statistical 
mechanics, represented by the MJ distribution, emerges as the $\kappa \to 0$ limit of the 
$\kappa$-generalized statistical mechanics. 
This fact reinforces the inclination to adopt the $\kappa$-generalized framework in quantum 
gravity phenomenology, wherein the flexibility offered by the $\kappa$ parameter broadens the 
scope of the relevant research domain. 

Furthermore, it should be noted that the MJ distribution maximizes the classical BGS entropy. 
Within the MJ approach, the classical energy-velocity relation is directly generalized to its 
relativistic counterpart. Consequently, this extension inherently fails to capture the power-law 
decay observed in the spectra of high-energy relativistic particles. By contrast, the non-extensive 
Kaniadakis entropy, given by Eq.~(\ref{eq:KS}), is rooted in an ab initio relativistic statistical 
formulation and explicitly preserves Lorentz symmetry, while simultaneously exhibiting power-law 
behavior at large momentum scales.

The theoretical relevance of the $\kappa$-generalized framework becomes particularly pronounced in the 
Planck regime, where the fundamental assumptions underlying conventional statistical mechanics may no 
longer hold. One possible avenue for incorporating quantum gravitational corrections arises from the 
postulates of DSR. Furthermore, the combination of $\kappa$-generalized statistical mechanics with DSR 
may offer a suitable theoretical framework at the Planck scale.

In the following sections, we employ this deformed formalism of statistical mechanics to systematically 
investigate the thermodynamic properties of a photon gas subject to Planck scale modifications, treating 
$\kappa$ as a phenomenological parameter that quantifies the deviations from standard photon gas 
thermodynamics.

\section{Photon Gas Thermodynamics at Planck-Scale with $\kappa$-generalized Statistical Mechanics}
\label{sec:kDSR}

We now investigate the photon gas thermodynamics within $\kappa$-generalized statistical mechanics 
(hereafter we refer to $\kappa$-DSR). To establish a unified framework that simultaneously accounts for 
both kinematic and statistical Planck 
scale modifications, we extend the standard statistical mechanics formulation to incorporate $\kappa$ 
parameter at Planck scale. 
As already noted in Sec.~\ref{sec:kapsm}, this procedure involves introducing the $\kappa$-deformed 
logarithmic and exponential functions, i.e., Eq.-s~\eqref{eq:kap_ln} and \eqref{eq:kap_exp}.
 
In $\kappa$-generalized statistical mechanics, photon gas is described by a deformed Bose-Einstein 
distribution~\cite{Kania2001a,Kania2001b}, depending on one continuous parameter $\kappa$ and is 
given by:
\beq
 f_{\kappa} (\epsilon) = \frac{1}{\exp_{\kappa} (\beta \epsilon) - 1} \, , 
\eeq
where $\kappa$-deformed exponential function is given 
by Eq.~\eqref{eq:kap_exp}: $\exp_{\kappa}(x) = \left( \sqrt{1 + \kappa^2 x^2} 
+ \kappa x \right)^{1/\kappa}$.

The number of photons with energy $\epsilon =\hbar \omega$ within a frequency range between $\omega$ 
and $\omega+d\omega$ is given by
\beq
 dN^{\kappa} = \frac{V}{\pi^2 c^3} \frac{\omega^2}{\exp_{\kappa} 
 \left(\frac{\hbar \omega}{k_{\rm B} T}\right) - 1} \, d\omega ~, 
\eeq
with a corresponding internal energy
\beq
\label{eq:dUdwk}
 dU^{\kappa} = \frac{V}{\pi^2 c^3} \frac{\hbar \omega^3}{\exp_{\kappa} 
 \left(\frac{\hbar \omega}{k_{\rm B} T}\right) - 1} \,  d\omega ~.
\eeq
Eq.~\eqref{eq:dUdwk} is the photon gas radiation law generalized in the framework of Kaniadakis 
$\kappa$ statistics. In the limit $\kappa \to 0$, it reduces 
to the usual Planck radiation law. In the limit of low frequencies $\hbar \omega \ll k_{\rm B} T$, 
the denominator 
in Eq.~\eqref{eq:dUdwk} lead to the classical Rayleigh-Jeans law, 
$dU = \frac{V}{\pi^2 c^3} k_{\rm B} T \omega^2 d\omega$. For $\hbar \omega \gg k_{\rm B} T$, 
Eq.~\eqref{eq:dUdwk} becomes  
\beq
 dU^{\kappa} = \frac{V}{\pi^2 c^3} \hbar\omega^3 \exp_{\kappa} \left( -\frac{\hbar \omega}{k_{\rm B} T} \right) 
 d\omega ~,
\eeq 
which can be viewed as a generalization of Wien's law.  

Integrating Eq.~\eqref{eq:dUdwk} over the frequencies up to the Planck scale $\Lambda$, we obtain the 
following generalized total energy,  
\beq
\label{eq:Ukap1}
 U^{(\kappa \rm DSR)} = \frac{V}{\pi^2 c^3} \int_0^{\Lambda} \frac{\hbar \omega^3}{\exp_{\kappa} 
 \left( \frac{\hbar \omega}{k_{\rm B} T} \right) - 1} \, d\omega~.
\eeq  

According to the standard procedure, we introduce the dimensionless variable 
$x = \beta \hbar \omega \equiv \frac{\hbar \omega}{k_{\rm B} T}$, 
Eq.~\eqref{eq:Ukap1} then becomes  
\beq
 U^{(\kappa \rm DSR)} = \frac{V}{\pi^2 \hbar^3 c^3} \left( k_{\rm B} T \right)^4 
 \int_0^{\beta \Lambda} \frac{x^3}{\exp_{\kappa} (x) - 1} \, dx ~.
\eeq  

For convenience, we introduce the function $J_{\kappa}^n (y)$~\cite{Alia2003}
\beq
\label{eq:Jkn}
 J_{\kappa}^n (y) = \int_0^y \frac{x^n}{\exp_{\kappa} (x) - 1} \, dx ~.
\eeq
Note that the integral in Eq.~\eqref{eq:Jkn} converges when $-1 < n < 1/|\kappa| -1$. This convergence 
condition may 
further restrict the admissible values of $\kappa$ for bosons or fermions.

With function $J_{\kappa}^n (y)$ we finally obtain
\beq
\label{eq:UkapJ}
 U^{(\kappa \rm DSR)} = \frac{V}{\pi^2 \hbar^3 c^3} \left( k_{\rm B} T \right)^4 J_{\kappa}^3 (\Lambda/k_{\rm B} T) ~.
\eeq

Note that in the limiting case $\kappa \to 0$ and $\beta \Lambda \to \infty$, all the above expressions 
reduce to their standard 
counterparts. In particular, the internal energy $U^{(\kappa \rm DSR)}$ reads as
\beq
 U_{\kappa \to 0, \Lambda \to \infty} = U^{\rm (SR)} = \frac{\pi^2 V}{15 \hbar^3 c^3} (k_{\rm B} T)^4 ~,
\eeq
which is expected well-known result in SR. When only the $\kappa \to 0$ limit is taken 
(without $\Lambda \to \infty$ limit), one restores the internal energy expression within DSR, 
$U^{\rm (DSR)}$. 

The generalized heat of a photon gas $C_{\rm V}^{\rm (\kappa DSR)}$ follows immediately by 
using the definition 
$C_{\rm V}^{\rm (\kappa DSR)} \equiv \left(\frac{\partial U^{\rm (\kappa DSR)}}{\partial T}\right)_{\rm V}$.
After some algebraic steps, $C_{\rm V}^{\rm (\kappa DSR)}$ reads
\beq
 C_{\rm V}^{\rm (\kappa DSR)} = \frac{V }{\pi^2 \hbar^3 c^3} \left[ 4 k_{\rm B}^4 T^3 \int_0^{\Lambda/k_{\rm B} T} 
\frac{x^3}{\exp_{\kappa}(x)-1} \, dx - \frac{\Lambda^4}{k_{\rm B} T} \frac{1}{\exp_{\kappa}({\Lambda/k_{\rm B} T}) - 1} \right] ~,
\eeq
and in terms of the function $J_{\kappa}^n$, it is
\beq
\label{eq:CVkdsr}
 C_{\rm V}^{\rm (\kappa DSR)} = \frac{V}{\pi^2 \hbar^3 c^3} \left[ 4 k_{\rm B}^4 T^3 
 J_{\kappa}^3 \left(\frac{\Lambda}{k_{\rm B} T} \right) 
 - \frac{\Lambda^4}{k_{\rm B} T} \frac{1}{\exp_{\kappa}(\Lambda/k_{\rm B} T) - 1} \right] ~.
\eeq
It can verified that in the $\kappa \to 0 $ and $\Lambda \to \infty$ limit, Eq.~\eqref{eq:CVkdsr} goes to the
well-known regular photon gas specific heat capacity $C_{\rm V}^{\rm (SR)}$, i.e.,
\beq
 \lim_{\kappa \to 0, \Lambda \to \infty} C_{\rm V}^{\rm (\kappa DSR)} = C_{\rm V}^{\rm (SR)} 
 = \frac{4 \pi^2 V }{15 \hbar^3 c^3} k_{\rm B}^4 T^3 ~,
\eeq
as expected.

Within the framework of the Kaniadakis statistics, the $\kappa$-deformed Helmholtz free energy is given by
\beq
\label{eq:Fkap}
 F^{(\kappa \rm DSR)} = -k_{\rm B} T \ln \Xi^{\rm (\kappa DSR)} ~,
\eeq
where $\Xi^{\rm (\kappa DSR)}$ is the $\kappa$-generalized partition function of the system. 
The corresponding internal energy $U^{(\kappa \rm DSR)}$ reads as
\beq
\label{eq:UkapX}
 U^{(\kappa \rm DSR)} = -\frac{\partial}{\partial \beta}\ln \Xi^{\rm (\kappa DSR)} ~.
\eeq

Combining Eq.~\eqref{eq:Fkap} and Eq.~\eqref{eq:UkapX}, and applying the derivative $dT= -k_{\rm B}T^{2}d\beta$, 
hence $\partial/\partial \beta = -k_{\rm B} T^2 \partial/\partial T$, one can construct the connection between 
the Helmholtz free energy and the internal energy,
\beq
\label{eq:FUint}
 F^{(\kappa \rm DSR)} = -T \int_0^{T} \frac{U^{(\kappa \rm DSR)}}{T'^2} \, dT' ~.
\eeq

Substituting Eq.~\eqref{eq:UkapJ} into Eq.~\eqref{eq:FUint} yields the expression for the $\kappa$-deformed 
Helmholtz free energy. We relegate the detailed derivation to Appendix~\ref{app:B} (see in particular 
Eq.~\eqref{eq:HelmFE}), here we present only the final result:
\beq
\label{eq:HFE2}
 F^{\rm (\kappa DSR)} = -\frac{V}{3 \pi^2 \hbar^3 c^3} \left[ (k_{\rm B} T)^4 J_{\kappa}^3(\Lambda/k_{\rm B} T)
 + \Lambda^3 k_{\rm B} T \int_{\Lambda/k_{\rm B} T}^\infty \frac{dy}{\exp_{\kappa}(y) - 1} \right] ~. 
\eeq

The $\kappa$-generalized pressure $P^{(\kappa \rm DSR)}$ and the entropy 
$S^{(\kappa \rm DSR)}$ can be obtained by taking the derivative of free energy $F^{(\kappa \rm DSR)}$ with 
respect to the volume $V$ and the temperature $T$, respectively,
\beq
 P^{(\kappa \rm DSR)} = -\left(\frac{\partial F^{(\kappa \rm DSR)}}{\partial V}\right)_{\rm T} ~, 
 \quad S^{(\kappa \rm DSR)} = -\left(\frac{\partial F^{(\kappa \rm DSR)}}{\partial T}\right)_{V} ~. \label{eq:PSkD}
\eeq

To get pressure and entropy, we define a function $\Phi_{\kappa} (y)$:
\beq
\label{eq:Phik}
 \Phi_{\kappa} (y) = \int_y^\infty \frac{dx}{\exp_{\kappa}(x) - 1} ~.
\eeq
It is worth noting that, in the limit $\kappa \to 0$, the function $\Phi_\kappa(y)$ reduces to 
$\Phi_0(y) \sim e^{-y}$, which exhibits the exponential decay. For large values of $x$, the 
$\kappa$-deformed exponential function admits the asymptotic behavior 
$\exp_\kappa(x) \sim (2\kappa x)^{-1/\kappa}$. As a result, the asymptotic form of $\Phi_\kappa(y)$ 
is given by
\beq
 \Phi_{\kappa} (y) \sim \int_y^\infty (2\kappa x)^{-1/\kappa} \, dx \propto y^{1-1/\kappa} ~,
\eeq
Thus, for a fixed deformation parameter $\kappa$, say $\kappa = 0.1$, the decay of $\Phi_\kappa(y)$ 
follows a power-law, specifically $\Phi_{\kappa=0.1}(y) \sim y^{-9}$. As $\kappa$ increases, 
the power-law decay becomes progressively slower. This asymptotic behavior will prove useful in 
interpreting the high-energy tails of the thermodynamic quantities examined in the subsequent sections.

The free energy in Eq.~\eqref{eq:HFE2} can now be rewritten as
\beq
\label{eq:HFEre}
 F^{\rm (\kappa DSR)} = -\frac{V}{3 \pi^2 \hbar^3 c^3} \left[ (k_{\rm B} T)^4 J_{\kappa}^3 (\Lambda/k_{\rm B} T)
  + \Lambda^3 k_{\rm B} T \Phi_{\kappa} (\Lambda/k_{\rm B} T)\right] ~.
\eeq
It is evident that the free energy receives contributions from two distinct terms: the first term, 
$(k_{\rm B} T)^4 J_{\kappa}^3 (\Lambda/k_{\rm B} T)$ (up to an overall constant prefactor), may be 
identified as the ``bulk'' contribution, while the second term, 
$\Lambda^3 k_{\rm B} T \Phi_{\kappa} (\Lambda/k_{\rm B} T)$, corresponds to the ``boundary'' contribution.
Both contributions will subsequently affect other thermodynamic quantities through partial derivatives, 
as will be demonstrated in the following analysis.

Recalling the expression of internal energy $U^{\rm (\kappa DSR)}$, Eq.~\eqref{eq:UkapJ}, one immediately 
finds
\beq
\label{eq:F-Uk}
 F^{\rm (\kappa DSR)} = -\frac{1}{3} U^{\rm (\kappa DSR)} (T) 
 - \frac{V}{3 \pi^2 \hbar^3 c^3} \Lambda^3 k_{\rm B} T \Phi_\kappa(\Lambda/k_{\rm B} T) ~.
\eeq

The pressure can be obtained according to Eq.~\eqref{eq:PSkD},  
\beq
\label{eq:PkD2}
 P^{\rm (\kappa DSR)} = \frac{1}{3 \pi^2 \hbar^3 c^3} \left[ (k_{\rm B} T)^4 J_{\kappa}^3 (\Lambda/k_{\rm B} T)
  + \Lambda^3 k_{\rm B} T \Phi_{\kappa} (\Lambda/k_{\rm B} T)\right] ~.
\eeq
Once again, the bulk and boundary contributions to the pressure are clearly identifiable. In particular, 
by making use of the expression for the internal energy $U^{\rm (\kappa DSR)} (T)$, the familiar 
\textit{universal} relation between pressure and internal energy (for a photon gas) is now modified by 
the boundary contribution term as follows:
\beq
\label{eq:u31p}
 P^{\rm (\kappa DSR)} = \frac{U^{\rm (\kappa DSR)}}{3V} 
 + \frac{\Lambda^3 k_{\rm B} T}{3 \pi^2 \hbar^3 c^3} \Phi_{\kappa} (\Lambda/k_{\rm B} T) ~.
\eeq
It is evident that the boundary contribution term $\Phi_{\kappa} (\Lambda/k_{\rm B} T)$ now affects the 
equation of state.

At this point, a noteworthy observation is in order. It is well known that for a photon gas, the pressure 
$P$ and internal energy $U$ satisfy the relation $P = U/3V$ (equivalently $P = u/3$, where $u \equiv U/V$ 
is the internal energy density). However, it is clear that this relation no longer holds in the $\kappa$-DSR 
framework. In fact, the extra term in Eq.~\eqref{eq:u31p}, namely 
$\frac{\Lambda^3 k_{\rm B} T}{3 \pi^2 \hbar^3 c^3} \Phi_{\kappa} (\Lambda/k_{\rm B} T)$, indicates that 
$\kappa$-DSR effects introduce an additional contribution to the pressure of the photon gas. This extra 
pressure term may also have observable implications in extreme astrophysical environments.

The entropy of the photon gas can be obtained analogously from Eq.~\eqref{eq:PSkD}, yielding
\beq
 S^{\rm (\kappa DSR)} = \frac{V }{3 \pi^2 \hbar^3 c^3} \left[ 4k_{\rm B}^4 T^3 J_{\kappa}^3 (\Lambda/k_{\rm B} T) 
 +\Lambda^3 k_{\rm B} \Phi_{\kappa} (\Lambda/k_{\rm B} T) \right] ~.
\eeq
As in the case of the free energy, this entropy also receives both bulk and boundary contributions. 
Nevertheless, in the appropriate limits $\Lambda \to \infty$ and $\kappa \to 0$, the entropy 
$S^{\rm (\kappa DSR)}$ reduces to the standard Stefan-Boltzmann (SB) entropy for blackbody radiation.

It is straightforward to verify that the total radiation energy can be expressed in the following form:
\beq
 U^{(\kappa \rm DSR)} = F^{(\kappa \rm DSR)} + TS^{(\kappa \rm DSR)} ~.
\eeq

The thermodynamic quantities of the photon gas presented above depend on the value of $\kappa$, and their 
evaluation requires numerical integration, as will be discussed in the next section.

\subsection{Numerical Results and Discussions}

We now proceed to numerically investigate the thermodynamic behavior of a photon gas within the 
frameworks of SR, DSR, and the $\kappa$-DSR formulation, respectively, building upon the analytical 
expressions derived in the preceding section. Throughout this section, we adopt natural units 
$\hbar =c=k_{\rm B}=G=1$.

In all numerical analyses, we set the Planck energy cutoff to $\Lambda=1$. The deformation 
parameter $\kappa$ is taken to be $\kappa =0.10$ and $\kappa = 0.22$. 
Note that the agreement between theoretical predictions and phenomenological implications of Kaniadakis'
formulation gives $\kappa = 0.2165$ (over a certain energy range) in high energy particle 
physics~\cite{Kania2002a}, which lies within the range of our parameter choices.
Temperatures are expressed in units of the Planck temperature, and the plotted range is restricted
to $0< T \leq 1$.

\begin{figure}[h!]
\centering
\includegraphics[width=0.7\linewidth]{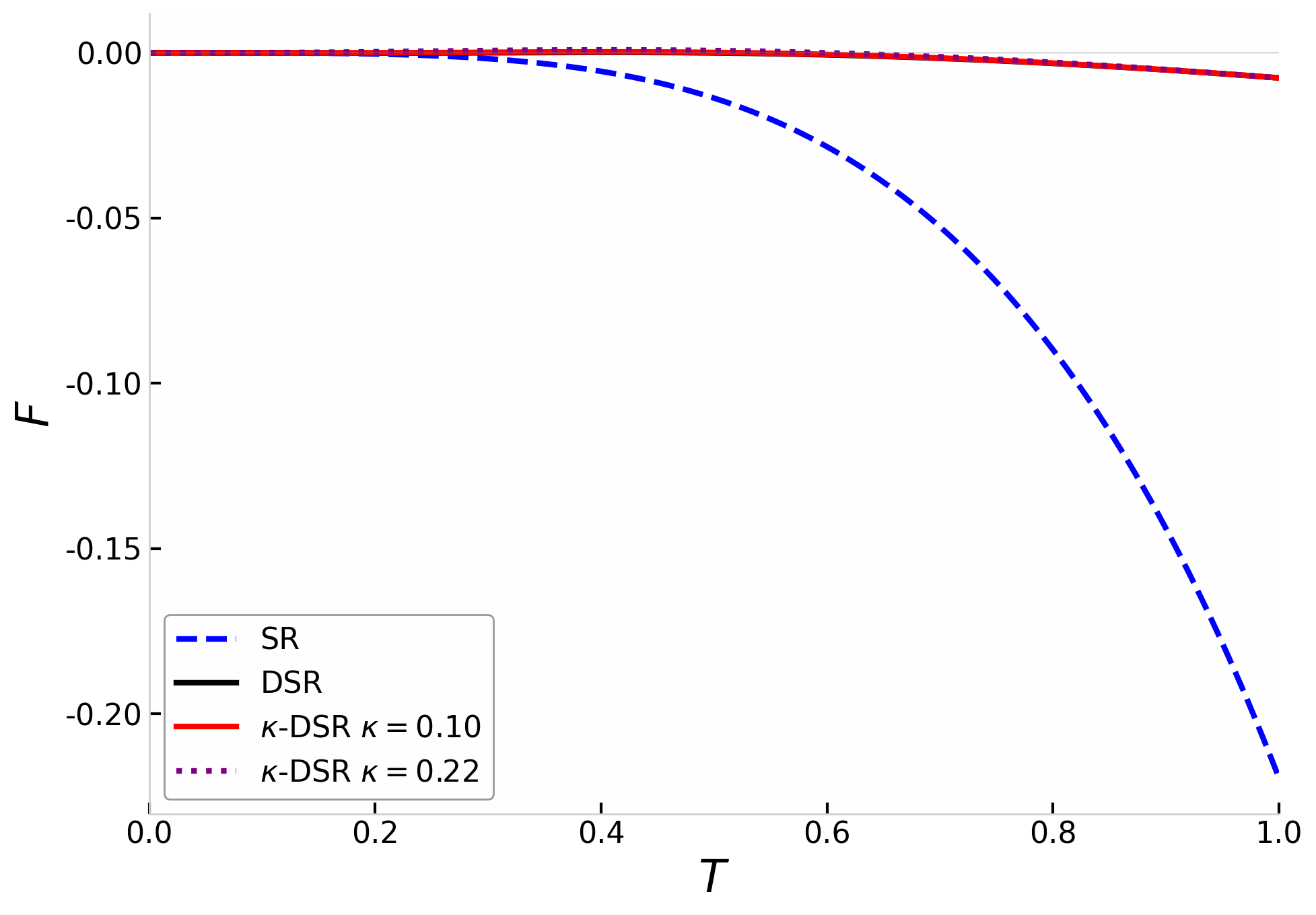}
 \caption {(Color online) The Helmholtz free energy $F$ is plotted as a function of the temperature 
 $T$ for the SR (blue dashed line), DSR (black solid line), and $\kappa$-DSR frameworks, with the 
 deformation parameter set to $\kappa = 0.10$ (red solid line) and $\kappa = 0.22$ 
 (purple dotted line), respectively.}
\label{fig:f-t}
\end{figure}

In Fig.~\ref{fig:f-t}, we plot the Helmholtz free energy $F$ as a function of the temperature $T$ for 
the SR (blue dashed line), DSR (black solid line), and $\kappa$-DSR (red solid line) frameworks, 
with the deformation parameter set to $\kappa = 0.10$. For comparison, we also include the free energy 
within the $\kappa$-DSR framework for $\kappa = 0.22$ (purple dotted line). 

As expected for a photon gas, the free energy decreases monotonically with increasing temperature across 
all three frameworks. However, within both the DSR and $\kappa$-DSR frameworks, the free energy is 
significantly suppressed at high temperatures compared to that in SR. This suppression is attributable 
to the presence of an upper cutoff imposed by the Planck energy scale $\Lambda$, which strongly restricts 
the accessible phase space and hence the number of microscopic states available to the photon gas in the 
high temperature regime.

On the other hand, the $\kappa$-DSR free energy exhibits larger values relative to the DSR one for both 
values of $\kappa$ considered in the present study, namely $\kappa = 0.10$ (red solid line) and 
$\kappa = 0.22$ (purple dotted line). This behavior simply follows from the fact that, for any allowed 
value of $\kappa$, the inequality $\exp_{\kappa}(x) < e^x$ always holds, which in turn implies 
$\frac{1}{\exp_{\kappa}(x)-1} > \frac{1}{e^{x}-1}$ for the corresponding statistical distribution factor. 
Consequently, the effective statistical weight is enhanced compared to that in the DSR case, which explains 
why the $\kappa$-DSR free energy curve always lies above the DSR one.

As already emphasized in the previous section, within the $\kappa$-DSR framework the free energy 
$F^{\rm (\kappa DSR)}$ receives both bulk and boundary contributions. These contributions become manifest 
when the temperature drops to around $T \sim 0.3$. In this regime, the $\kappa$-DSR free energy not only 
succeeds the DSR free energy $F^{\rm (DSR)}$, but also exceeds the SR free energy $F^{\rm SR}$. 
The physical implications of this behavior will subsequently propagate to the equation of state (pressure) 
and the entropy, as will be shown below.

In Fig.~\ref{fig:u-t}, we present the internal energy $U$ as a function of temperature $T$, shown with 
both (a) a linear scale and (b) a log-log scale. In the linear-scale subplot, as the temperature approaches 
the Planck regime, the internal energy curve for the SR case rises steeply, following the standard 
radiation behavior. In contrast, the DSR curve exhibits a considerably slower increase relative to the 
SR case, a consequence of the finite energy cutoff $\Lambda$. The $\kappa$-DSR curve again interpolates 
between the SR and DSR scenarios, suggesting that the $\kappa$-deformed distribution enhances the 
thermal contribution of the accessible states compared to the pure DSR framework.

As the temperature decreases, the internal energy also monotonically decreases. In particular, in the 
low-temperature regime, roughly around $T \lesssim 0.3$, the three curves remain closely aligned and 
are difficult to distinguish in the linear-scale plot of Fig.~\ref{fig:u-t}(a). Their differences are 
more clearly visible in the log-log scale, as shown in Fig.~\ref{fig:u-t}(b). It is evident in the 
low-temperature region, the DSR internal energy reduces to the SR value, as expected. For the 
$\kappa$-DSR internal energy with $\kappa = 0.10$, the deviation from the DSR case is small, while for 
$\kappa = 0.22$, the deviation is quite apparent. Nevertheless, as $T \to 0$, all three internal 
energies vanish.

\begin{figure}[h!]
 \centering
 \includegraphics[width=0.7\linewidth]{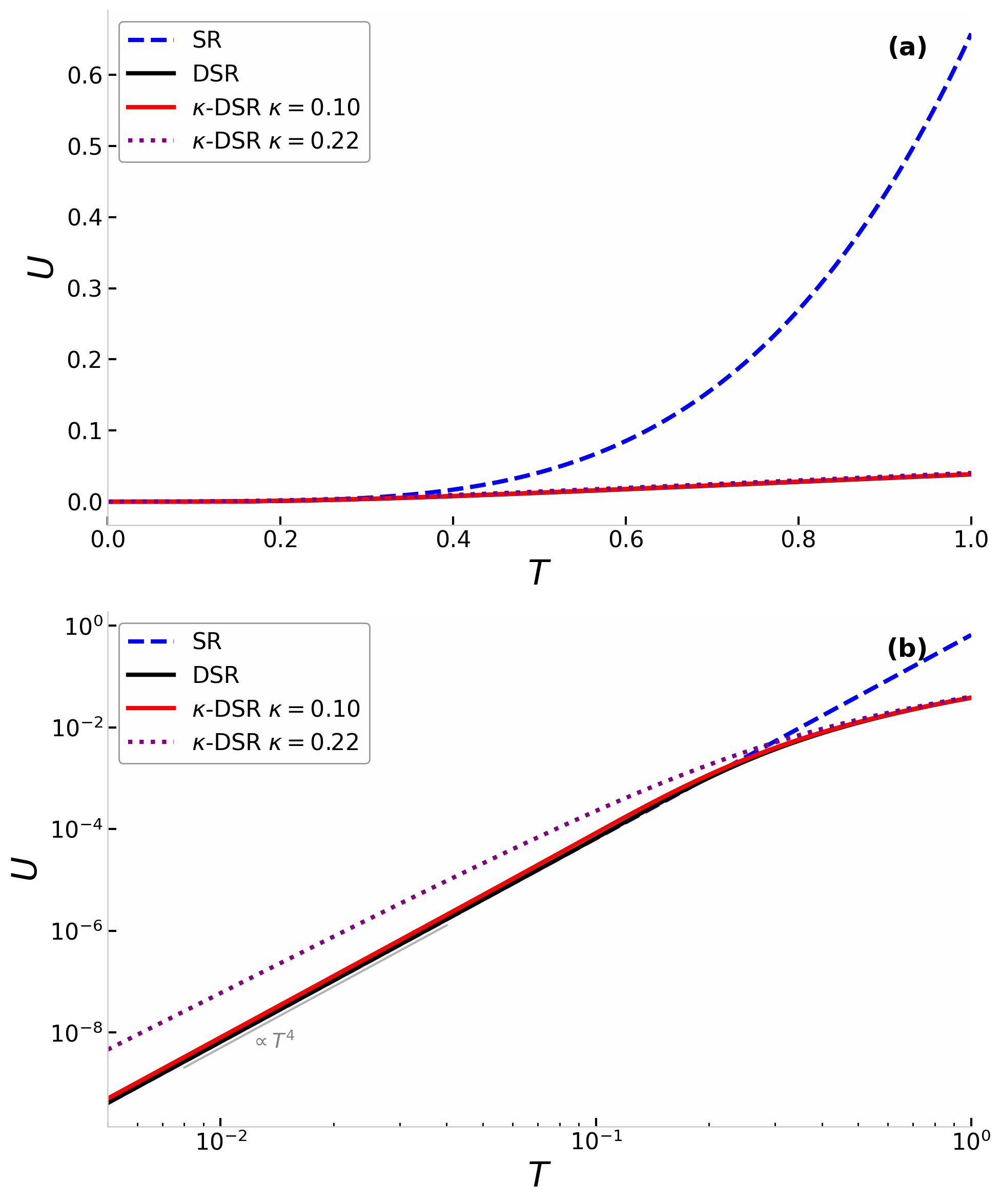}
 \caption{(Color online) The internal energy $U$ of a photon gas is plotted as a function of 
 temperature $T$ for the SR case (blue dashed line), the DSR case (black solid line), and the 
 $\kappa$-DSR cases with the deformation parameter set to $\kappa = 0.10$ (red solid line) and 
 $\kappa = 0.22$ (purple dotted line).} 
 \label{fig:u-t}
\end{figure}

\begin{figure}[h!]
 \centering
 \includegraphics[width=0.7\linewidth]{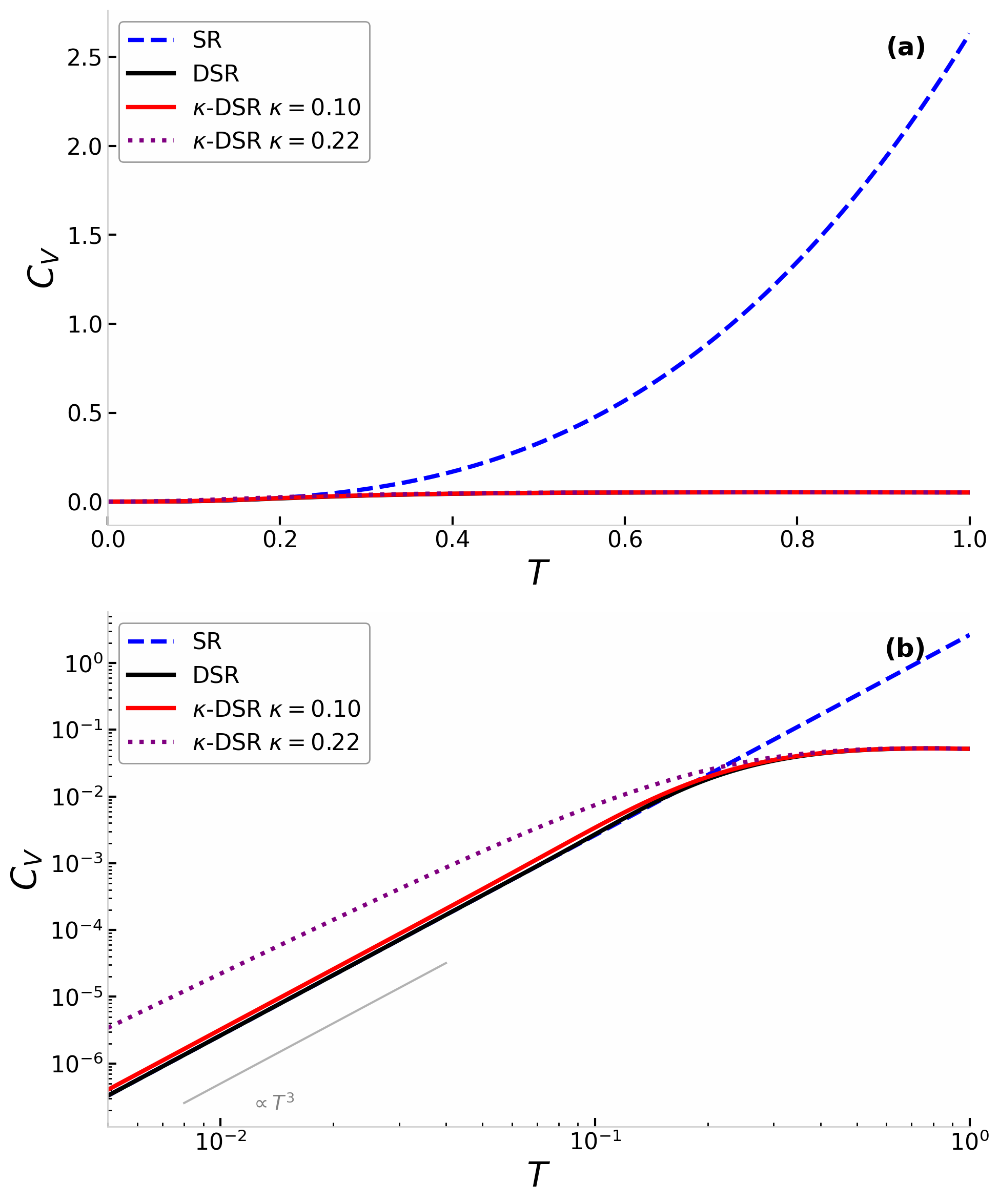}
 \caption{(Color online) Following the same layout as Fig.~\ref{fig:u-t}, the heat capacity 
 $C_{\rm V}$ of a photon gas is presented as a function of temperature $T$ for the DSR case 
 (black solid line) and the $\kappa$-DSR cases with $\kappa = 0.10$ (red solid line) and 
 $\kappa = 0.22$ (purple dotted line), respectively, using (a) a linear scale and (b) a log-log 
 scale. The log-log scale better distinguishes between the different values of $\kappa$.} 
 \label{fig:cv-t}
\end{figure}

Similarly, in Fig.~\ref{fig:cv-t}, we present the heat capacity $C_{\rm V}$ of a photon gas as 
a function of temperature $T$, shown with (a) a linear scale and (b) a log-log scale, for the 
SR case (blue dashed line), the DSR case (black solid line), and the $\kappa$-DSR cases with 
$\kappa = 0.10$ (red solid line) and $\kappa = 0.22$ (purple dotted line). Recall that in the 
SR case, $C_{\rm V}^{\rm (SR)}$ follows the standard $T^3$ scaling, characteristic of an ordinary 
photon gas. In contrast, the DSR curve, denoted $C_{\rm V}^{(\rm DSR)}$, exhibits a clear 
suppression and tends toward a slower growth rate as the temperature approaches the Planck scale. 
This behavior reflects the fact that once the high-energy states near the cutoff become thermally 
populated, further increases in temperature can no longer activate an unlimited number of 
additional photon modes.

Generally speaking, the $\kappa$-DSR heat capacity $C_{\rm V}^{(\kappa\rm DSR)}$ exceeds the DSR 
value. However, as the temperature approaches the Planck scale, the $C_{\rm V}^{(\kappa\rm DSR)}$ 
curves for the chosen values of $\kappa$ asymptotically reduce to the DSR curve, as can be seen 
from the high-temperature end in Fig.~\ref{fig:cv-t}(b). Like the internal energy, when the 
temperature drops below $T \sim 0.3$, the $\kappa$-deformation enhances the thermal response of 
the photon gas. The log-log scale highlights the difference between the DSR and $\kappa$-DSR 
results, and it is evident that a larger value of $\kappa$ corresponds to a higher heat capacity 
curve. Meanwhile, note that both the DSR and $\kappa$-DSR heat capacities vanish as $T \to 0$, 
as expected from the third law of thermodynamics.

\begin{figure}[h!]
 \centering
 \includegraphics[width=0.7\linewidth]{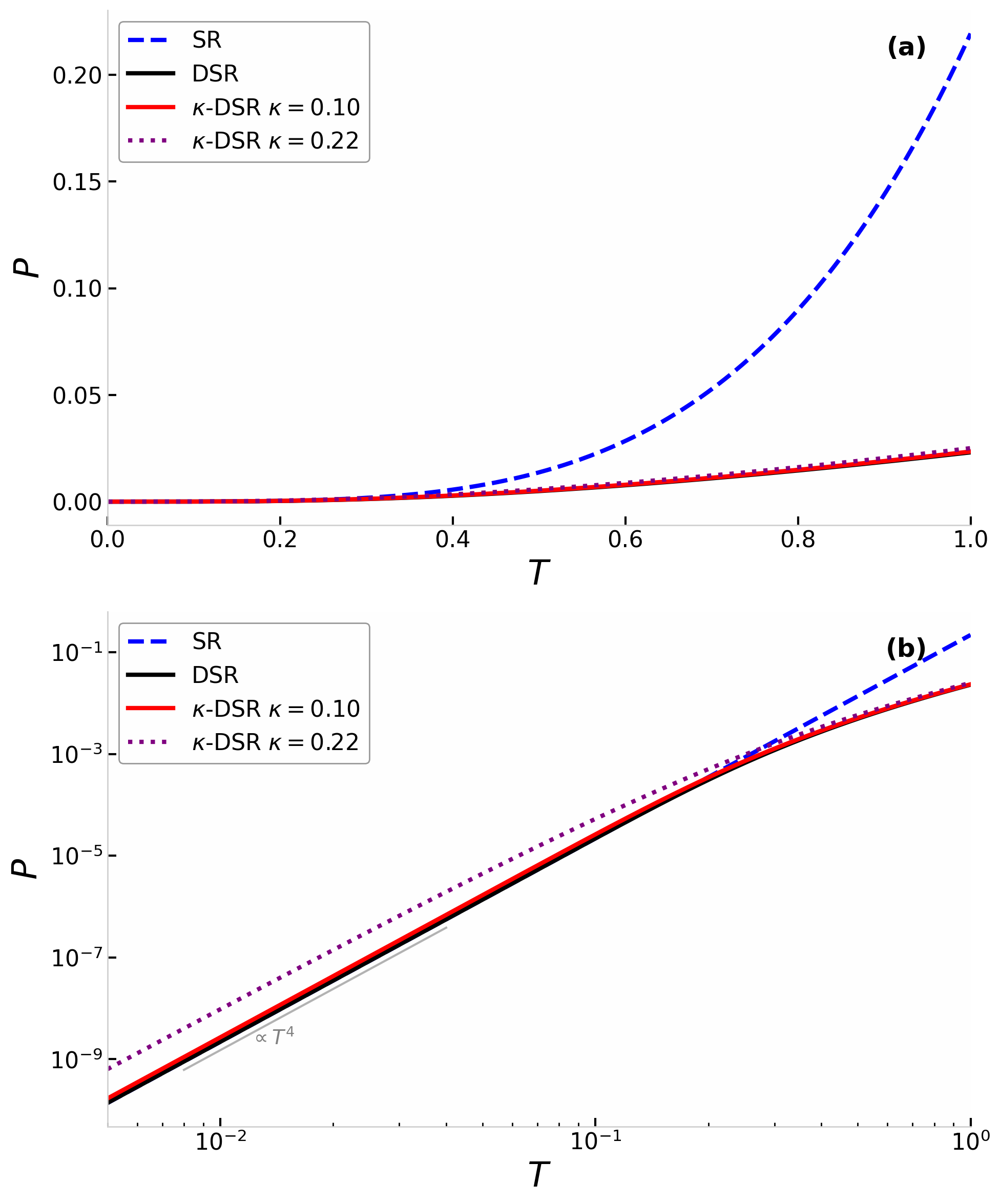}
 \caption{(Color online) Similar to Fig.~\ref{fig:u-t}, but for the pressure $P$ of a photon gas, 
 shown with (a) a linear scale and (b) a log-log scale. The curves correspond to the SR case (blue 
 dashed line), the DSR case (black solid line), and the $\kappa$-DSR cases with $\kappa = 0.10$ 
 (red solid line) and $\kappa = 0.22$ (purple dotted line).} 
 \label{fig:p-t}
\end{figure}

The pressure of a photon gas is plotted in Fig.~\ref{fig:p-t}, shown with (a) a linear scale 
and (b) a log-log scale. Its temperature dependence exhibits a trend closely resembling that of the 
internal energy. Similarly, compared to the SR pressure, both the DSR pressure $P^{(\rm DSR)}$ 
(black solid line) and the $\kappa$-DSR pressure $P^{(\kappa \rm DSR)}$ are strongly suppressed near 
the Planck scale, with the latter remaining consistently larger than the former for the two 
representative values of the deformation parameter considered, $\kappa = 0.10$ (red solid line) and 
$\kappa = 0.22$ (purple dotted line). This behavior confirms that the combined deformation modifies 
the equation of state through two entangled effects: the DSR cutoff reduces the accessible phase 
space, while the $\kappa$-statistical deformation alters the statistical weight assigned to high 
energy states below the cutoff.

This is consistent with the bulk and boundary contributions in Eq.~\eqref{eq:PkD2}. For instance, 
in the low-temperature regime, where $\Lambda/k_{\rm B} T \gg 1$, the boundary contribution arising 
from $\Phi_{\kappa}(\Lambda/k_{\rm B} T)$ becomes negligible, leaving only the bulk contribution. 
Consequently, the pressure in this region approximately satisfies a $T^4$ scaling, which originates 
from the bulk term via $T^4 J_{\kappa}^3(\Lambda/k_{\rm B} T) \propto T^4$.

\begin{figure}[h!]
 \centering
 \includegraphics[width=0.7\linewidth]{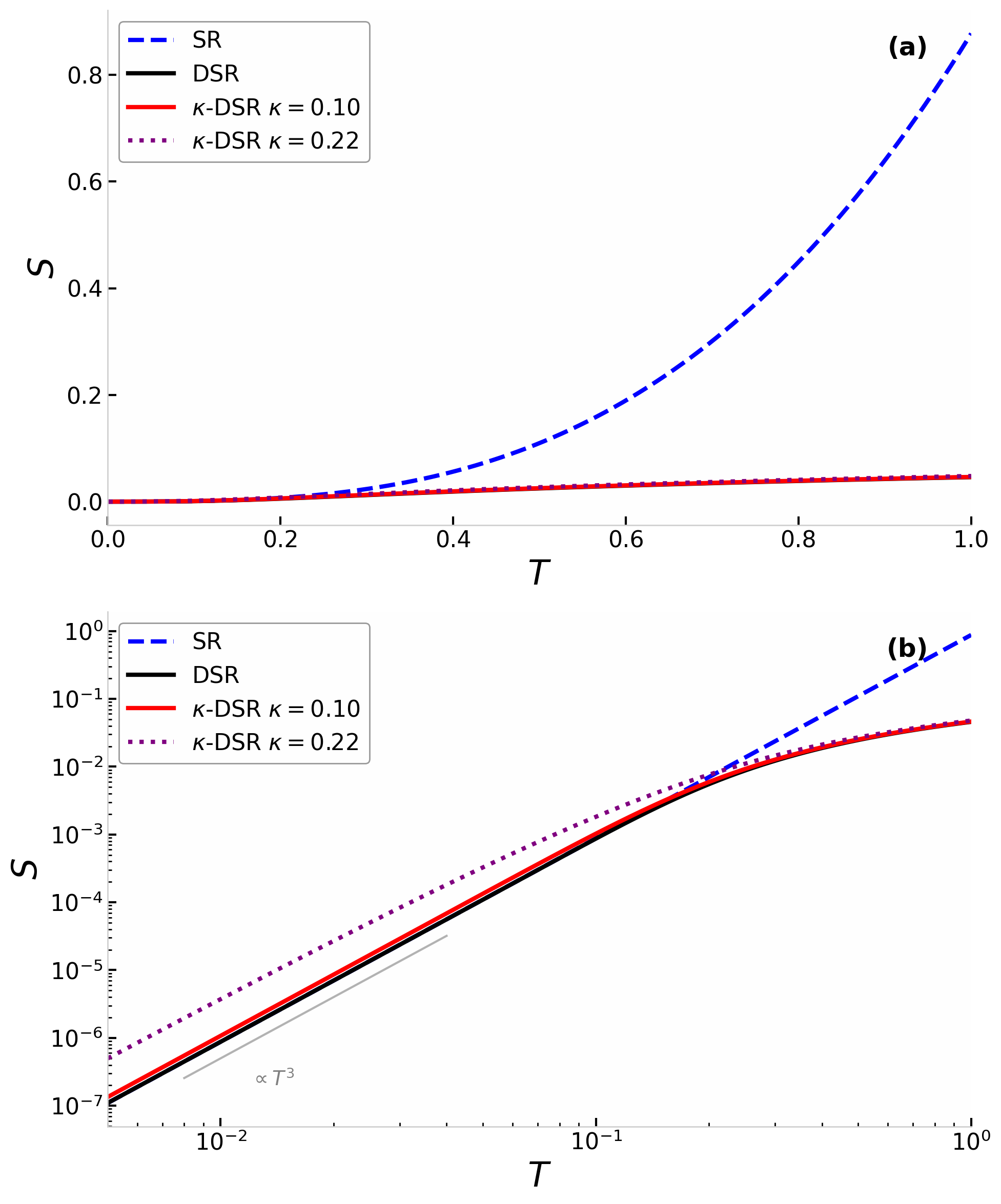}
 \caption{(Color online) Similar to Fig.~\ref{fig:u-t}, but for the entropy $S$ of a photon gas. 
 The curves correspond to the SR case (blue dashed line), the DSR case (black solid line), and 
 the $\kappa$-DSR cases with $\kappa = 0.10$ (red solid line) and $\kappa = 0.22$ (purple dotted line), 
 using (a) a linear scale and (b) a log-log scale.} 
 \label{fig:s-t}
\end{figure}

Finally, Fig.~\ref{fig:s-t} displays the entropy $S$ as a function of temperature $T$. In the SR case, 
$S^{\rm (SR)}$ (blue dashed line) follows the standard $T^3$ law of the SB entropy, as expected for a 
conventional photon gas thermodynamics. 
In contrast, the DSR entropy $S^{\rm (DSR)}$, due to the constraint imposed by the Planck energy scale 
$\Lambda$, is significantly smaller than its SR counterpart in the high temperature regime, reflecting 
the reduction in available phase space volume induced by the Planck scale upper cutoff. The $\kappa$-DSR 
entropy lies above the curve of the DSR entropy, suggesting that the $\kappa$-deformation enhances the 
effective number of thermally accessible states relative to the DSR case. 
However, when comparing the 
$\kappa$-DSR and SR entropies, one finds that as the temperature decreases to roughly $T \lesssim 0.2$, 
the former begins to exceed the latter. This intriguing behavior reflects an intrinsic feature of 
$\kappa$-deformed statistics: the deformation effectively \textit{increases} the number of accessible 
microstates in this regime, thereby allowing the entropy to surpass the standard SB value. 

To better illustrate this behavior, Fig.~\ref{fig:s-t} (b) represents the Fig.~\ref{fig:s-t}(a) in 
a log-log format. In the figure, the black dotted line corresponds to the standard SB entropy 
$S^{\rm (SR)}$, the black solid line to the DSR entropy $S^{\rm (DSR)}$, the red solid line to the 
$\kappa$-DSR entropy $S^{\rm (\kappa DSR)}$ with $\kappa = 0.10$, and the purple dotted line to the 
case with $\kappa = 0.22$. Below $T \sim 0.2$, the DSR entropy $S^{\rm (DSR)}$ already coincides with 
the SR entropy, as is clearly visible in the figure. In contrast, the $\kappa$-DSR entropy 
$S^{\rm (\kappa DSR)}$ begins to exceed both the DSR and SR entropies in this regime. 

\section{Summary and conclusions}
\label{sec:remk}

In this work, we investigate the thermodynamics of a photon gas at the Planck scale within the 
framework of doubly special relativity (DSR), formulated in terms of $\kappa$-deformed statistical 
mechanics. Our analysis is based on a careful combination of three key ingredients: the modified 
dispersion relation, which notably preserves Lorentz symmetry; the Planck-scale upper bound on 
accessible modes; and the $\kappa$-generalized statistical mechanical framework. 
The $\kappa$-deformed thermodynamic 
quantities of the photon gas, including the internal energy $U$, Helmholtz free energy $F$, 
heat capacity $C_{\rm V}$, pressure $P$, and entropy $S$, are subsequently derived. 
We highlight the nontrivial effects induced by the deformation parameter $\kappa$ on these 
thermodynamic quantities, in particular on the entropy $S$ and on the experimentally relevant 
observable, e.g., the heat capacity $C_{\rm V}$ and the pressure $P$ (equation of state). 
The corresponding $\kappa$-dependent contributions are explicitly expressed.

We present numerical results illustrating the temperature dependence of these thermodynamic 
quantities for the SR, DSR, and $\kappa$-DSR scenarios, taking two representative values of the 
deformation parameter, $\kappa = 0.10$ and $0.22$. As the temperature approaches the Planck scale, 
the $\kappa$-DSR results deviate significantly from the standard SR predictions. In the SR case, 
the available phase space is unbounded, and the thermodynamic quantities follow the usual 
Stefan-Boltzmann scaling. By contrast, both the $\kappa$-DSR and DSR curves are strongly suppressed, 
a suppression originating from the Planck scale upper bound $\Lambda$, which removes the contribution 
of arbitrarily high energy photon modes. The additional $\kappa$-deformation modifies the statistical 
weights of the allowed photon states and partially compensates for this DSR suppression.

On the other hand, when the temperature decreases to around $T \lesssim 0.2$, the $\kappa$-deformed 
statistical distribution, governed by the factor $(\exp_{\kappa}(x) -1)^{-1}$, plays an important 
role in the thermodynamic quantities in this low temperature regime. In particular, it leads to a 
larger $\kappa$-DSR heat capacity, pressure, and entropy, compared to the SR one. Moreover, 
the universal relation $P = u/3$ (where $u$ is the energy density), which holds in SR, no longer 
persists in the $\kappa$-DSR framework; instead, an extra $\kappa$-dependent term appears in the 
equation of state. These nontrivial features provide useful phenomenological signatures of the 
$\kappa$-DSR framework and 
motivate a systematic scan over the parameter $\kappa$, which would clarify the sensitivity of 
thermodynamic observable to the strength of the generalized statistical corrections.

The present framework may also serve as a useful point of departure for several extensions. First, 
our results suggest that photon gas thermodynamics provides a viable theoretical setting for probing 
Planck scale modifications. In this context, the DSR-imposed upper bound governs the ultraviolet 
accessibility of photon states, whereas the $\kappa$-deformation controls their effective statistical 
occupation. This separation offers a useful conceptual basis for investigating generalized radiation 
thermodynamics in early universe and/or black-hole scenarios.

Second, the combined corrections arising from DSR and $\kappa$-statistics constitute a thermodynamic 
probe of Planck scale phenomenology, applicable not only to photon gas thermodynamics, but more 
generally to massive bosons and fermions. For massive bosons, the present framework can be directly 
generalized to the general $\kappa$-deformed Bose-Einstein statistics with non-vanishing chemical 
potential, in contrast to the zero chemical potential considered in current study~\cite{YZhang2026}. 
This more general scenario has some interesting connections with black hole thermodynamics, see, e.g.,
Ref.~\cite{Wald1997,Aref2025}.
For fermions, modifications to the Fermi energy level 
induced by $\kappa$-related effects may lead to further corrections to the Chandrasekhar limit for 
white dwarf stars, as well as in other astrophysical contexts.

Recent observations of ultra-high-energy (UHE) photons from gamma-ray bursts (GRBs), reported by the 
High-Altitude Water Cherenkov Observatory (HAWC), the Tibet AS$_\gamma$ experiment, and the Large High 
Altitude Air Shower Observatory (LHAASO) collaborations, offer a sensitive testbed for Planck scale 
modifications to Lorentz symmetry~\cite{HAWC2020,AS2021,LA2021}.
In particular, the highest detected photon energies, together with the approximate power-law behavior 
of the overall spectrum (across energies spanning from $10^{10}$ eV to $10^{20}$ eV), may challenge 
conventional expectations derived from standard dispersion relations. These developments motivate a 
systematic theoretical exploration of modified dispersion relations and their phenomenological 
implications within the framework of $\kappa$-deformed doubly special relativity ($\kappa$-DSR), which 
we defer to future investigations.

\section{Acknowledgement}

The authors thank J. Zhang for helpful discussions. This work is partially supported by the National 
Natural Science Foundation of China (NSFC) via Grant No. 11505115. 

\section*{Appendices}

\appendix

\section{Brief Summary of a Photon Gas Thermodynamics within DSR Framework}
\label{app:A}

In this appendix, we derive the photon gas thermodynamics within the framework of DSR and summarize 
the principal  thermodynamic quantities of a photon gas. Although the main results of 
this derivation have been previously reported in Ref.~\cite{Xiong2026}, 
we reproduce them here for the sake of self-contained completeness. We subsequently list the 
principal DSR-modified thermodynamic quantities of the photon gas.

The ordinary thermodynamic properties of a photon gas are well-established in 
standard thermodynamics and statistical physics, see, e.g.~\cite{Landau1980}. Before exploring the 
thermodynamic behaviors 
in DSR with the $\kappa$-generalized statistical mechanics framework, we briefly
summarize the results of the photon gas thermodynamics in DSR framework. 

It is a well-known fact that photon gas obeys Bose-Einstein statistics with zero chemical potential 
$\mu=0$, $f(\epsilon) = \frac{1}{e^{\beta \epsilon} - 1}$, where $\beta=1/k_{\rm B} T$. 
The number of available 
microstates in a volume $V$, within the momentum range from $p$ to $p+dp$ is given by 
$\frac{V}{\pi^2 \hbar^3} p^2 dp$. 
The corresponding partition function is
\beq
 \ln \Xi = -\int_0^{\infty} g(\epsilon) \, 
 \ln\left(1 - e^{-\beta \epsilon}\right) \, d\epsilon ~, \label{eq:lnXi}
\eeq
where $g(\epsilon)$ is the density of states is $g(\epsilon) = \frac{V}{\pi^2 \hbar^3 c^3} \epsilon^2$.

When calculating the thermodynamic properties of a photon gas using the MS model within DSR, 
the above expression, Eq.~\eqref{eq:lnXi}, must be modified, while $g(\epsilon)$ remains the 
same as in standard statistical mechanics. This is because the modified dispersion relation, 
Eq.~\eqref{eq:mdisp}, for 
massless particles such as photons does not deviate from the usual SR, and still preserves 
the relation $E = p$~\cite{Ameli2001}. In DSR, in addition to the speed of light $c$, a 
second invariant quantity, the Planck energy $\Lambda$, is introduced, implying the existence 
of an upper limit of energy 
on the order of the Planck energy scale. Owing to the presence of $\Lambda$, the above equation 
Eq.~\eqref{eq:lnXi} is modified to the following form:
\beq
 \ln \Xi^{\rm (DSR)} = -\int_0^{\Lambda} g(\epsilon) \, 
 \ln\left(1 - e^{-\beta \epsilon}\right) \, d\epsilon ~. \label{eq:lnXdsr}
\eeq
Note that within the framework of DSR, the upper limit of the integral is required by the 
second invariant quantity introduced by DSR.  
It is evident that since the photon gas satisfies $m = 0$, the dispersion relation remains 
unchanged and consistent with that of SR. However, the partition function is modified due 
to the presence of the upper limit $\Lambda$, and in the limit $\Lambda \to \infty$, 
the special relativistic form is recovered.

With the specific form of log-partition function $\ln \Xi$ in Eq.~\eqref{eq:lnXdsr}, according 
to the standard statistical mechanics, all the thermodynamic quantities can be derived. 

Upon introducing the dimensionless variable $x = \beta \epsilon$,
and then integrating by parts, we obtain
\beq
 \ln \Xi^{\rm (DSR)} = -\frac{V}{3 \pi^2 \hbar^3 c^3} \left[\Lambda^3 \ln \left(1 - e^{-\beta \Lambda}\right) 
        - \frac{1}{\beta^3}\int_{0}^{\beta \Lambda} \frac{x^3}{e^x - 1} \, dx \right] ~.
\eeq

The Helmholtz free energy follows from the standard relation
$F^{\rm (DSR)}  = -\beta^{-1} \ln \Xi^{\rm (DSR)}$, namely
\beq
  F^{\rm (DSR)} = \frac{V}{3 \pi^2 \hbar^3 c^3} \left[ \frac{\Lambda^3}{\beta} \ln(1 - e^{-\beta \Lambda}) 
  - \frac{1}{\beta^4}\int_{0}^{\beta \Lambda} \frac{x^3}{e^x - 1} \, dx \right] ~.
\eeq

The total internal energy is obtained from $U^{\rm (DSR)} = 
-\frac{\partial}{\partial \beta} \ln \Xi^{\rm (DSR)}$, 
which gives
\beq
  U^{\rm (DSR)}= \frac{V}{\pi^2 \hbar^3 c^3 \beta^4} \int_{0}^{\beta \Lambda} \frac{x^3}{e^x - 1} \, dx ~.
\eeq

The specific heat at constant volume can be obtained from $C_{\rm V} = (\frac{\partial U}{\partial T})_{\rm V}$, 
yielding
\beq
 C_{\rm V}^{\rm (DSR)} = \frac{V}{\pi^2 \hbar^3 c^3} \left[ \frac{4}{\beta^3} 
 \int_0^{\beta \Lambda} \frac{x^3}{e^x - 1} \, dx 
 - \frac{\beta \Lambda^4}{e^{\beta \Lambda} - 1}\right] ~, \label{eq:CVdsr}
\eeq
or, explicitly expressed in the temperature $T$:
\beq
 C_{\rm V}^{\rm (DSR)} = \frac{V}{\pi^2 \hbar^3 c^3} \left[ 4(k_{\rm B} T)^3 \int_0^{\Lambda/k_{\rm B} T} \frac{x^3}{e^x - 1} \, dx 
- \frac{\Lambda^4/k_{\rm B}T}{e^{\Lambda/k_{\rm B}T} - 1}\right] ~. \label{eq:CVdsrT}
\eeq
Note that the contribution from the first term in Eq.~(\ref{eq:CVdsr}) is constrained by the 
existence of an upper limit set by the Planck energy scale. Evidently, when the Planck energy 
scale $\Lambda \to \infty$ in DSR at some finite temperature, 
so that $\Lambda \to \infty$, the integral in the first term reduces to the standard 
Bose-Einstein integral, 
i.e., $\int_{0}^{\infty} \frac{x^3}{e^x - 1} \, dx = \Gamma(4)\zeta(4) = \frac{\pi^4}{15}$. 
In this limit, the second term is governed by a Bose-Einstein type distribution factor, 
specifically the 
$e^{\Lambda/k_{\rm B} T}$  term in the denominator, and consequently vanishes as $\Lambda/T \to \infty$.
Thus, the thermodynamic description of the photon gas smoothly reduces to the standard SR 
results, e.g., 
$F^{\rm (SR)} = -\frac{\pi^2 V}{45\hbar^3 c^3} (k_{\rm B} T)^4$, and the heat capacity 
$C_{\rm V}^{\rm (SR)} = \frac{4\pi^2 V}{15\hbar^3 c^3} (k_{\rm B} T)^3$, as can be readily verified.

Using the thermodynamic definition $P = -(\frac{\partial F}{\partial V})_{\rm T}$, and noting 
that the 
free energy is proportional to the volume, the pressure is thus
\beq
 P^{\rm (DSR)} = \frac{1}{3 \pi^2 \hbar^3 c^3} 
              \left[ \frac{1}{\beta^4} \int_0^{\beta \Lambda} \frac{x^3}{e^x - 1} dx 
              - \frac{\Lambda^3}{\beta} \ln(1 - e^{-\beta \Lambda}) \right] ~.
\eeq

Similarly, the entropy $S$ follows from the thermodynamic identity, $S = (U-F)/T$, leading to
\beq
 S^{\rm (DSR)} = \frac{V}{3 \pi^2 \hbar^3 c^3} \left[ \frac{4}{\beta^3} 
 \int_0^{\beta \Lambda} \frac{x^3}{e^x - 1} \, dx - \Lambda^3 \ln(1 - e^{-\beta \Lambda}) \right] ~.
\eeq

\section{Derive Photon Gas Helmholtz Free Energy within $\kappa$-DSR Framework}
\label{app:B}

In this appendix we first derive the Helmholtz free energy of a photon gas within the 
$\kappa$-DSR framework. The corresponding $\kappa$-generalized 
thermodynamic quantities such as the pressure $P^{\rm (\kappa DSR)}$, and/or the entropy
$S^{\rm (\kappa DSR)}$ can subsequently be obtained by applying this free energy expression, 
as discussed in the main text (see Sec.~\ref{sec:kDSR} for details).

The Helmholtz free energy within $\kappa$ statistical mechanics $F^{\rm (\kappa DSR)}$ can be 
determined by similar procedure to that whthin standard statistical mechanics:
\beq
\label{eq:FTprim}
 F^{\rm (\kappa DSR)} (T) = -T \int_0^T \frac{U^{\rm (\kappa DSR)}(T')}{T'^2} \, dT' ~,
\eeq
with
\beq
 U^{\rm (\kappa DSR)} (T) = \frac{V}{\pi^2 \hbar^3 c^3} (k_{\rm B} T)^4 J_{\kappa}^3\left(\frac{\Lambda}{k_{\rm B} T} \right) ~,
\eeq
where $J_{\kappa}^n (y)$ is defined in Eq.~\eqref{eq:Jkn}) in Sec. IV:
\beq
 J_{\kappa}^n (y) = \int_0^{y} \frac{x^3}{\exp_{\kappa}(x) - 1} \, dx ~. \nonumber
\eeq
For convenience, let us write the Eq.~\eqref{eq:FTprim} in a more compact form, with the help of Eq.~\eqref{eq:Jkn}
\beq
 F^{\rm (\kappa DSR)} (T) = -\mathcal{A} T \int_0^T T'^2 J_{\kappa}^3 
 \left(\frac{\Lambda}{k_{\rm B} T'} \right)\, dT' ~,
\eeq
where $\mathcal{A} = \frac{V}{\pi^2 \hbar^3c^3}$.
After performing the integration over $T'$, we have the expression of free energy,
\beq
\label{eq:Fdouble}
 F^{\rm (\kappa DSR)} (T) = -\frac{V \Lambda^3}{\pi^2 \hbar^3 c^3} (k_{\rm B} T)
 \int_{\Lambda/k_{\rm B} T}^{\infty} \frac{1}{y^4} 
 \left[ \int_0^y \frac{x^3}{\exp_{\kappa}(x) - 1} \, dx \right] dy ~.
\eeq

We may further reduce the double integral in Eq.~\eqref{eq:Fdouble} to a single integral.
Let
\beq
\label{eq:Intdef}
 \mathcal{I} (T) = \int_{\Lambda/k_{\rm B} T}^{\infty} \frac{1}{y^4} 
\left[ \int_0^y \frac{x^3}{\exp_{\kappa}(x) - 1} \, dx \right] dy ~.
\eeq

We compute $\mathcal{I}(T)$ via using integration by parts, then after some steps we arrive at
\beq
 \mathcal{I} (T) = \frac{1}{3} \left(\frac{k_{\rm B} T}{\Lambda}\right)^3 J_{\kappa}^3(\Lambda/k_{\rm B} T)
 + \frac{1}{3} \int_{\Lambda/k_{\rm B} T}^\infty \frac{dy}{\exp_{\kappa}(y) - 1} ~.
\label{eq:Isingle}
\eeq 
Substituting Eq.~\eqref{eq:Isingle} $\mathcal{I} (T)$ back into $F^{\rm (\kappa DSR)}(T)$, finally we have
\beq
\label{eq:HelmFE}
 F^{\rm (\kappa DSR)} = -\frac{V}{3 \pi^2 \hbar^3 c^3} \left[ (k_{\rm B} T)^4 J_{\kappa}^3(\Lambda/k_{\rm B} T)
 + \Lambda^3 k_{\rm B} T \int_{\Lambda/k_{\rm B} T}^\infty \frac{dy}{\exp_{\kappa}(y) - 1} \right] ~.
\eeq

Note that Eq.~\eqref{eq:HelmFE}, in the limit $\kappa \to 0$, and further of $\Lambda \to \infty$, 
it can be eventually verified 
\beq
 \lim_{\kappa \to 0, \Lambda \to \infty} F^{\rm (\kappa DSR)} = -\frac{\pi^2 V (k_{\rm B} T)^4}{45 \hbar^3 c^3} ~,
\eeq
which is the standard free energy of a photon gas, as expected.

The expression of free energy $F^{\rm (\kappa DSR)}$ will be further used to produce thermodynamic
quantities such as pressure $P^{\rm (\kappa DSR)}$ and entropy $S^{\rm (\kappa DSR)}$, as we show 
in the main text.


\end{document}